\documentclass[epj]{svjour}
\usepackage{graphicx}
\usepackage{amsmath,amssymb}

\begin{document}
\title{Electron transport across a quantum wire in the presence of
electron leakage to a substrate}

\author{
     Tomasz Kwapi\'nski  \inst{1,2} \thanks{Electronic address:
     tomasz.kwapinski@umcs.lublin.pl}
\and Sigmund Kohler\inst{3} \and Peter H{\"{a}}nggi\inst{1} }
\institute{Institut f\"ur Physik, Universit\"at Augsburg,
Univerist\"atsstr.~1, 86135 Augsburg, Germany \and Instytut
Fizyki, Uniwersytet Marii Curie-Sk\l odowskiej, 20-031 Lublin,
Poland \and Instituto de Ciencia de Materiales de Madrid, CSIC,
Cantoblanco, 28049 Madrid, Spain}
%
\date{Date: \today}
\abstract{We investigate electron transport through a mono-atomic wire
which is tunnel coupled to two electrodes and also to the underlying
substrate.  The setup is modeled by a tight-binding Hamiltonian and
can be realized with a scanning tunnel microscope (STM).  The
transmission of the wire is obtained from the corresponding Green's
function.  If the wire is scanned by the contacting STM tip, the
conductance as a function of the tip position exhibits oscillations
which may change significantly upon increasing the number of wire atoms.  Our
numerical studies reveal that the conductance depends strongly on
whether or not the substrate electrons are localized.  As a further
ubiquitous feature, we observe the formation of charge oscillations.
} 
\maketitle
\section{\label{sec1}Introduction}

\begin{figure}[b]
\centerline{\includegraphics[width=.5\columnwidth]{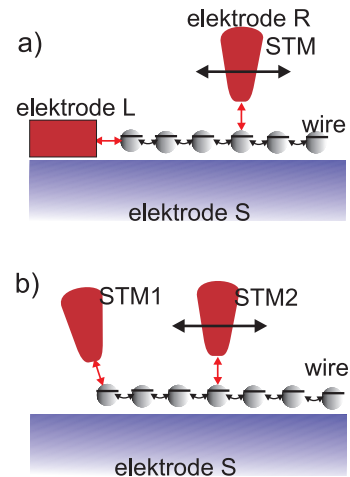}}
\caption{Schematic view of a quantum wire on a surface (electrode
S) contacted at sites $1$ and $4$.  The contacts is established by
(a) one fixed electrode and one STM tip or (b) two STM tips. In
both configurations, the right electrode can be moved.}
\label{Fig1}
\end{figure}

Mono-atomic wires of metal atoms fabricated on a surface are the
ultimately small conductors and may be used in nanoelectronics to
connect nanodevices such as quantum gates, qubits, or
nanotransistors. Thus their electronic properties are of crucial
interest.  One-dimensional mono-atomic wires can be fabricated using
mechanically controlled break junctions \cite{5,6}. Such wires
are freely suspended and, thus, do not interact with any substrate.
For the same reason, they are somewhat unstable, and it
consequently represents a challenge to form long wires. Similar but
more stable structures can be fabricated on vicinal surfaces and
investigated with scanning tunneling microscopes (STM) \cite{1,2,3},
see Fig.~\ref{Fig1}. Well ordered and even longer examples are
double stranded gold wires grown on silicon vicinal surfaces such as
Si(335) and Si(557) \cite{2,3,4}. The geometry and the electronic
structure of these setups are sufficiently stable such that
measurements can be repeated many times. Notice that STM experiments
mainly focus on the electron transport from the STM tip to the
surface (perpendicular transport) and, thus, are not concerned with
the transport from one end of the wire to the other end.  The latter
type of STM experiments would require some modifications of the
setup as is discussed in Refs.~\cite{28,Tegen,Tanik,Wang0}.

The conductance of ideal or disturbed wires has been investigated
both experimentally and theoretically.  For example, it has been
predicted that the conductance of an atom chain depends on whether
the number of atoms is even or odd \cite{7,8,9,10,11}.  These
even-odd oscillations have been confirmed experimentally \cite{5}.
Conductance oscillations with larger periods may occur as well
\cite{12,13,14,15}.  They stem from a Fabry-Perot like resonance of
electrons with Fermi wavelength in the chain, which eventually changes the
chain filling factor \cite{12,14,15}. Moreover, the formation of
charge waves inside a wire was observed for both, non-magnetic
\cite{16} and magnetic wires \cite{17}.  The fact that the resonance
condition depends on the presence of further atoms beyond the
contact also leaves its fingerprints in the conductance, where
interference effects can be observed \cite{18,19,20,21,22,23}.

In this paper we consider setups in which one lead is realized by a
movable STM tip by which an atom of choice can be contacted; see
Fig.~\ref{Fig1}.  We describe the atom chain by a tight-binding
Hamiltonian and obtain the conductance within a Green's function
approach \cite{11,14,24,25,26,27,HRY,KohlerPR}.  It will turn out that the
conductance oscillations are influenced by small leakage currents from
the wire to the substrate.  Our substrate model, the spatial
separation of the wire atoms is taken into account by considering
local tunnel couplings.  A relevant parameter of this model is the
Fermi wavelength of the substrate electrons, which allows one to
interpolate between two limits: In the one limit, each wire atom couples
individually to a substrate with localized electrons, while for the
other limit model, the substrate represents a reservoir of delocalized
electrons.  These two setups may be interpreted as an insulating and
a conducting substrate, respectively, or the coupling to a molecule
with according localization properties \cite{Petrov}.

The paper is organized as follows. In Sec.~\ref{sec2} we present our
model and a scattering formalism.  Moreover, we derive an analytical
expression for the retarded Green's function of a wire coupled to a
surface.  With these expressions at hand, we investigate in
Sec.~\ref{sec:conductance} both conductance oscillations and charge
oscillations.  The main conclusions are drawn in
Sec.~\ref{sec:conclusions}.

\section{\label{sec2} Theoretical model and formalism }

We consider the setups sketched in Fig.~\ref{Fig1} which both consist
of a quantum wire connected to two metallic electrodes.  The wire may
exchange electrons also with the surface which, thus, represents a
further, weakly connected electrode.  One of these electrodes may be
fabricated by epitaxy or grown on the surface and is fixed
\cite{Wang0}.  The other electrode is movable and contacts an
atom of choice.  Alternatively, both electrodes may be realized by an
STM tip as is sketched in Fig.~\ref{Fig1}(b).

The model Hamiltonian for a wire with $N$ atom sites can be
written in the form $H = H_{0} + H_\mathrm{tun}$, where
\begin{equation}
\label{H0}
H_{0} = \sum_{\vec k\alpha=L,R,S} \varepsilon_{\vec k\alpha}
a^+_{\vec k\alpha} a_{\vec k\alpha} + \sum_{i=1}^N \varepsilon_i
a^+_i a_i
\end{equation}
describes the electrons in the wire and in the leads.  Electron
transitions between the leads and the wire are established by the
tunnel Hamiltonian
\begin{equation}
\begin{split}
H_\mathrm{tun}
={} & \sum_{\vec k} V_{\vec kL} a^+_{\vec kL}a_{m}
    + \sum_{\vec k} V_{\vec kR} a^+_{\vec kR}a_{n}
\\&
    + \sum_{i=1}^{N-1} V_i a^+_i a_{i+1}
    + \text{h.c.}
\end{split}
\label{eq2}
\end{equation}
Here, $m$ and $n$ label the atom connected to the left and to the
right lead, respectively.  It is worth mentioning that if the left
STM electrode couples to the first wire site, $m=1$ and the right
electrode to the last site, $n=N$, the system corresponds to the
break junction geometry of Refs.~\cite{5,9,10,11,14}.  The
operators $a_i$ and $a^+_i$ create and annihilate, respectively,
an electron at site $i=1,\ldots,N$, while $a_{\vec k\alpha}$ and
$a^+_{\vec k\alpha}$ are the according leads operators.  The
tunnel matrix elements $V_{{\vec k}\ell}$ enter the expressions
for the current only via the spectral densities
$\Gamma^{\ell}=2\pi \sum_{\vec k}|V_{\vec k\ell}|^2
\delta(\varepsilon-\varepsilon_{\vec k\ell})$, $\ell=L,R$, which
we model within a wide-band approximation as energy independent.
With the above Hamiltonian we assume that electron-electron
interactions do not lead to correlation effects and can be
captured by an effective shift of the onsite energies. Then both
spin directions are independent of each other, such that the spin
need not be considered explicitly. For Au or Pb chains on vicinal
silicon surfaces, these conditions are met reasonably well.
Generally, this should hold for non-magnetic wire atoms \cite{31}.
It has also been shown that electron-electron correlations do not
change the period of conductance oscillations \cite{15,29,30}.

In order to describe electron leakage from the wire to the surface, we
consider the surface as a further, weakly coupled electrode.  Thus, we
introduce the wire-surface tunneling Hamiltonian
\begin{equation}
H_\text{w-s}
= \sum_{\vec k} V_{i\vec k}^s a^+_{\vec k}a_{i} + \text{h.c.},
\label{Hws}
\end{equation}
where $V_{j\vec k}^s = V_{\vec k}^s \exp(i\vec k\vec R_j)$ is the
tunnel matrix element for atom $j$ \cite{32,33}.  The phase factor
reflects the position of atom $j$ and has the consequence that the
leakage depends on the spatial separation of the atoms.  Assuming
equal distances $a$ between neighboring atoms, while mainly substrate
electrons with Fermi wavelength play a role, we obtain $V_{j\vec k}^s
= V_{\vec k}^s \exp(ik_F ja)$.  As for the leads, the influence of the
substrate can be subsumed in a spectral density.  It will turn out
that the relevant quantity reads
\begin{equation}
\Gamma^S_{ij}= \Gamma^S \frac{\sin(k_F a|i-j|)}{k_F a|i-j|} \;,
\end{equation}
with the effective leakage strength $\Gamma^S$.
A formally similar coupling and spectral density has been used to
describe decoherence of spatially separated qubits coupled to a
bosonic environment \cite{Palma,DollEPL,DollPRB}.  While the fermionic
case can still be treated within scattering theory, the bosonic model
gives rise to memory effects which may be considered within a
non-Markovian master equation approach
\cite{Kleinekathofer,Welack,Weiss}

Two limiting cases are worth being discussed:
(i) If $k_F a\gg 1$, the spectral density is rather small unless
$i=j$.  This means that we can employ the approximation
$\Gamma^S_{ij}=\Gamma^S \delta_{ij}$.  This describes a substrate
with a very short mean free path such as a semi-conductor or an
insulator. Then an electron that tunnels from a particular atom to
the substrate can re-enter only at the same site. Obviously, this
scheme corresponds to a model in which each wire atom is coupled to
an individual additional electrode.
(ii) In the opposite limit, $k_Fa\ll 1$, we obtain
$\Gamma^S_{ij}=\Gamma^S$.  Physically this means that the wire
electrons tunnel to a delocalized substrate orbital, as is the
case for metallic surfaces.

In order to obtain the linear conductance between the electrodes
$L$ and $R$ and the local density of states at site $i$, $\rho_i$,
one needs to compute the Green's function for the total
Hamiltonian. The linear conductance at zero temperature is given
by the Landauer formula \cite{24,25,26,27,KohlerPR}
\begin{eqnarray}\label{cond}
G = \frac{e^2}{h}T(E_F)
= \frac{2e^2}{h} \Gamma^L \Gamma^R |G^r_{m n}(N,E_F)|^2 ,
\end{eqnarray}
where $T(E_F)$ is the electron transmission at the Fermi energy which
we choose to be $E_F=0$.  Using the equation of motion for the
retarded Green's function $G^r_{mn}$, one finds the elements $G^r_{m
n}$ from the relation $G^r_{mn}=(\hat A^{-1})_{mn}$,
where the matrix $\hat A$ is given by
\begin{equation}
\begin{split}
(\hat A)_{ij}
={} &(\varepsilon-\varepsilon_i)\delta_{i,j}-V_i(\delta_{i,j+1}+\delta_{i+1,j})
\\
& +i\frac{\Gamma^L}{2} \delta_{i,m}\delta_{m,j}+i\frac{\Gamma^R}{2}
\delta_{i,n}\delta_{n,j} +i\frac{\Gamma^S_{ij}}{2} \;.
\end{split}
\end{equation}
Below, when presenting specific results, we will always assume that
both leads couple equally strong to the wire, yielding  $\Gamma^L
=\Gamma^R =\Gamma$, and that all onsite energies and intra-wire
tunnel matrix elements are position independent,
$\varepsilon_i=\varepsilon_0$ and $V_i=V$.  These assumptions are
quite reasonable for a wire consisting of one atom species in an
equidistant arrangement on the surface.

In the absence of wire-surface tunneling, i.e.\ for $\Gamma^S=0$,
the matrix $\hat A_{ij}$ becomes tri-diagonal and its inverse, the
Green's function, can be computed analytically.  After some
algebra we find
\begin{equation}
\label{G2}
\begin{split}
G^r_{m n}
={}& {(-V)^{n-m} \det{A_0^{N-n}} } \big\{ \det A_0^N+i\frac{\Gamma}{2}
     \Phi_{m,n}
\\
&-   \frac{\Gamma^2}{4}\det A_0^{m-1}\det A_0^{N-n} \det A_0^{n-m-1}
     \big\}^{-1} ,
\end{split}
\end{equation}
where
\begin{equation}
\Phi_{n1,n2}
=\det A_0^{n1-1} \det A_0^{N-n1}+\det A_0^{n2-1}\det A_0^{N-n2} ,
\end{equation}
while $A_0^N$ denotes the tri-diagonal $N\times N$ matrix for
$\Gamma^S = \Gamma = 0$, i.e.\ for the isolated wire. The determinant
of this matrix can be evaluated to read $\det A^N_0=V^N u_N(\phi)$,
where $u_N(\phi)$ is the $N$th Chebyshev polynomial of the second kind
and $\phi=\arccos\{(\varepsilon-\varepsilon_0/2V)\}$ plays the role of
a Bloch phase \cite{16}. Note that $\det A^0_0=1$ and $\det
A^1_0=\varepsilon-\varepsilon_0$.

For a wire-surface coupling in the limit (i), the additional
self-energy is proportional to the unit matrix, $\Gamma_{ij}^S =
\Gamma^S \delta_{ij}$, such that ${\hat A}^{-1}$ can still be
computed analytically.  Then we obtain again the Green's function
\eqref{G2}, but with $\varepsilon-\varepsilon_0$ replaced by
$\varepsilon-\varepsilon_0 -i\Gamma^S/2$.

The local density of states at wire site $i$ is determined by the
retarded Green's function $G^r_{ii}$ owing to the relation
$\rho_{i}(\varepsilon) = -\mathop{\mathrm{Im}} {
(\mathop{\mathrm{cof}}{\hat A^N_{ii}} / \pi \det{\hat A^N}})$, where
$\mathop{\mathrm{cof}}{\hat A^N_{ii}}$ denotes the algebraic
complement of the matrix ${\hat A^N_{ii}}$, the so-called cofactor.
The charge localized at site $i$ can be obtained by integrating the
local density of states up to the Fermi energy,
\begin{equation}
Q_i=\int^{E_F}_{-\infty} \rho_i (\varepsilon) d\varepsilon .
\end{equation}
Note, that due to wire-surface coupling, analytical formulas for the
local density of states or charge density $Q_i$ cannot be simplified
further.  Thus Eq.~\eqref{G2} is the most general analytical
expressions for the retarded Green functions for arbitrary wire
length.

\section{Conductance oscillations}
\label{sec:conductance}

The conductance of a quantum wire is governed by the electron wave
functions at the Fermi surface.  In particular for short wires, it
may even be such that a single orbital in the relevant energy
range dominates.  Its overlap with the electrodes may change
significantly with the length of the wire and, thus, the
conductance changes as well.  Typically this change appears as
periodic oscillation \cite{7,8,9,10,11,12,13,14,15}. For a break
junction geometry, i.e., when the left electrode is connected to
the first atom ($m=1$), and the right electrode to the last atom
($n=N$), the period $M$ of the oscillation can be determined
analytically: Writing the transmission $T(E_F)$ in terms of the
Green's function \eqref{G2}, one imposes that this transmission
for a wire of length $N$ must be the same as for a wire of length
$N+M$. This results in the condition $ \cos({\pi l /
M})=(E_F-\varepsilon_0)/2V$, where $l=1,...,M-1$ \cite{14}. The
same relation holds for a wire coupled to a surface.

\subsection{Quantum wire isolated from the substrate}

Before addressing the influence of a substrate, we study the
conductance oscillations of a wire that couples only to the
electrodes, but not to the substrate.  In particular, we focus on
the influence of the wire atoms beyond the STM tips. The presence
of these atoms modifies the wave functions of the wire electrons
and, consequently, it may affect the overlap between the relevant
wire states and the electrodes.  In all our numerical studies
presented below, we use the Fermi energy $E_F=0$ as reference
point and assume a fixed left electrode i.e.\ $m=1$. All energies
are measured in units of the wire-electrode coupling $\Gamma$,
such that formally $\Gamma^L=\Gamma^R= 1$, while the conductance
$G$ is plotted in units of the conductance quantum $G_0 = e^2/h$,
such that it becomes identical to the dimensionless transmission
$T(E_F)$.

\begin{figure}
\centerline{\includegraphics{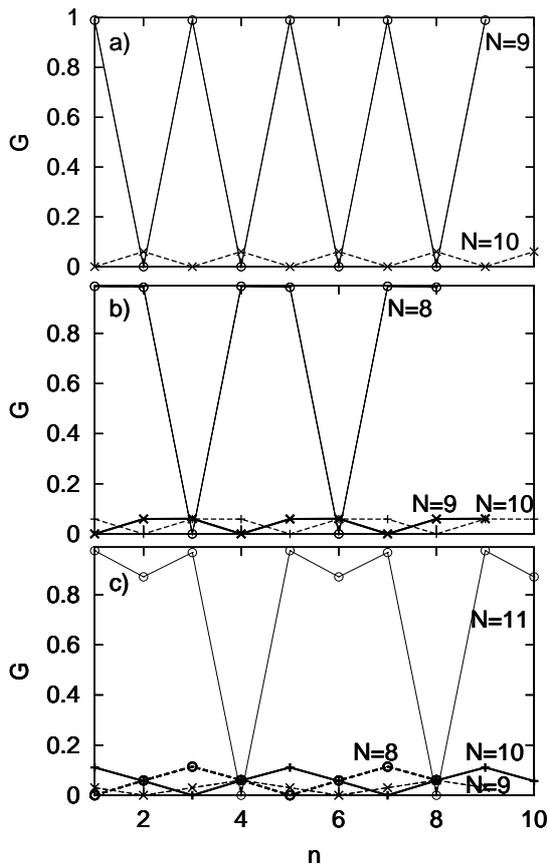}} \caption{\label{Fig2} The
conductance (in units of $[e^2/h]$ so that $G\doteq T$) as a
function of the STM tip position $n$ for the wire lengths
$N=8,9,10,11$ in the absence of wire-substrate coupling, i.e., for
$\Gamma^S=0$.  The onsite energies are (a) $\varepsilon_0 = 0$, (b)
$\varepsilon_0=V$, and (c) $\varepsilon_0=\sqrt{2}V$. These values
correspond to conductance oscillations with periods $M=2,3,4$.  The
intra-wire tunnel matrix elements is $V=4\Gamma$. The lines serve as
a guide to the eye.}
\end{figure}

Figure~\ref{Fig2} depicts the conductance of wires with various
lengths $N$ as a function of the tip position $n$; cf.\
Fig.~\ref{Fig1}(a). The onsite energies are position independent,
$\varepsilon_i = \varepsilon_0$, and chosen such that they satisfy
the oscillation condition for period $M=2,3,4$ (panels a, b, and c,
respectively).  A most prominent feature is that the amplitude of
the emerging oscillations depends strongly on the total wire length
$N$, or put differently, on the number of atoms beyond the right
electrode.  A closer look at the results for periodicity $M=2$
(panel a) reveals that the amplitude changes with each additional
wire atom from a large to a small value and back. For odd $N$, the
conductance oscillates with a very large amplitude, whereas for an
even number of atoms, these oscillations are very small. Note, that
the conductance obtained for any even $N$ is hardly distinguishable
form the one for $N=10$. The same holds for any odd $N$ and $N=9$.

For the periods $M=3$ and $M=4$, we find large
oscillation amplitudes for the wire lengths $N=5,8,11,\dots$ and
$N=3,7,11,\ldots$, respectively.  For other lengths, the
conductance still oscillates, but its value never exceeds 10\% of
the conductance quantum.  The bottom line of these numerical
investigations is that we observe strong oscillations with period
$M$ provided that the wire has length $N = Mk-1$, where
$k=1,2,3,\ldots$ is any natural number \cite{14}. Thus for
periods $M=2,3,4$ as considered in Fig.~\ref{Fig2}, maximal
amplitudes of the conductance oscillations are observed for
\begin{equation}
N = \begin{cases}
1,3,5,\ldots  & \text{for $M=2$,} \\
2,5,8,\ldots  & \text{for $M=3$,} \\
3,7,11,\ldots & \text{for $M=4$.}
\end{cases}
\end{equation}

\begin{figure}
\centerline{\includegraphics[width=.7\columnwidth]{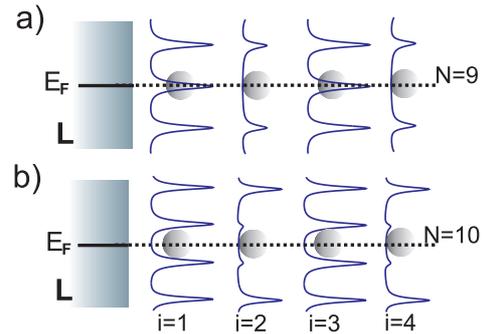}}
 \caption{\label{Fig3}
Local density of states at atom wire sites $i=1,2,3,4$ for wires
that consist of (a) $N=9$ and (b) $N=10$ atoms.  The onsite energies
are $\varepsilon_0=0$, such that the conductance obeys oscillations
with period $M=2$, i.e.\ even-odd oscillations.  The other
parameters are as in Fig.~\ref{Fig2}.  The horizontal lines mark the
Fermi energy.}
\end{figure}

Some insight about the physical origin of the conductance oscillations
is provided by considering the local density of states
$\rho_i(\varepsilon)$.   For a case with even-odd oscillations
($M=2$), this is shown in Fig.~\ref{Fig3} for the wire sites
$i=1,2,3,4$.  The magnitude of the conductance can be understood from
the fact that for $N=9$ (panel a), the density of states possesses a
peak at the Fermi energy.  Thus, electrons from the Fermi surface of
the left electrode can tunnel to the wire resonantly, which
facilitates transport.  For length $N=10$ (panel b), by contrast, the
density of states at the first site vanishes, such that transport is
practically blocked.  The density of states at the further sites
changes with period $M=2$.  Applying the same arguments to the right
STM tip explains that the conductance must oscillate with the same
period.  In the second case (even $N$), however, the value of the
conductance is small due to the small local density of states at the
Fermi energy on the left STM tip .  This directly translates to a
small oscillation amplitude of the conductance. In the same way one
can explain the conductance oscillations with other periods. In
general, the conductance is maximal when both STM electrodes are
connected to sites that possess a large local density of states at the
Fermi level.

A possible application of our results is the experimental estimate of
the onsite energy $\varepsilon_0$ which strongly influences the
oscillation periods.  However, since also the wire length $N$ and the
tip position $n$ have significant impact on the conductance, it would
be necessary to perform several measurement with wires that differ
only in length but are identical otherwise.

\subsection{Influence of the substrate}

\begin{figure}
\centerline{\includegraphics[width=.75\columnwidth]{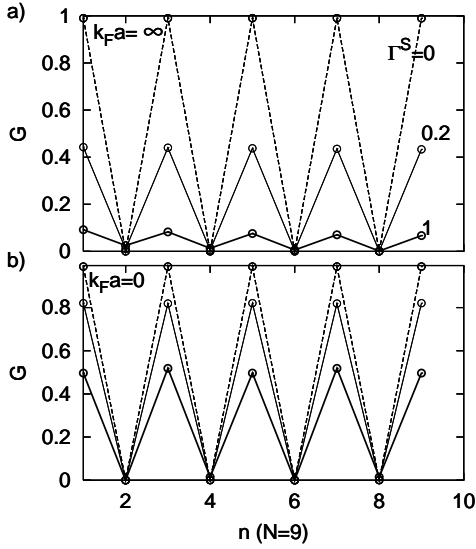}}
\caption{\label{Fig5} The conductance in units [$e^2/h$] as a
function of the STM position $n$ for a wire consisting of $N=9$
atoms with onsite energies $\varepsilon_0=0$.  The wire-surface
coupling strength is $\Gamma^S=0$ (dashed line),
$\Gamma_S=0.2\Gamma$ (thin solid), and $\Gamma$ (solid).  The other
parameters are as in Fig.~\ref{Fig2}. (a) Short wavelength limit
$k_Fa=\infty$, such that each atom couples to an individual
environment and $\Gamma^S_{ij}=\Gamma^S \delta_{ij}$. (b) $k_Fa=0$,
such that all atoms couple collectively to the substrate and, thus,
$\Gamma^S_{ij}=\Gamma^S$. }
\end{figure}

The influence of the wire-substrate tunneling can be appreciated
in Fig.~\ref{Fig5} where we compare the conductance oscillations
for two values of $\Gamma^S$ with those obtained in the absence of
the substrate, i.e.\ for $\Gamma^S=0$.  Let us first consider the
limiting cases in which the substrate electrons are perfectly
delocalized ($k_Fa=0$) or perfectly localized ($k_Fa=\infty$).  We
find two significant features: The oscillation period is
practically not influenced by the leakage, while the oscillation
amplitude decreases with increasing coupling strength.

\begin{figure}[t]
\centerline{\includegraphics[width=.95\columnwidth]{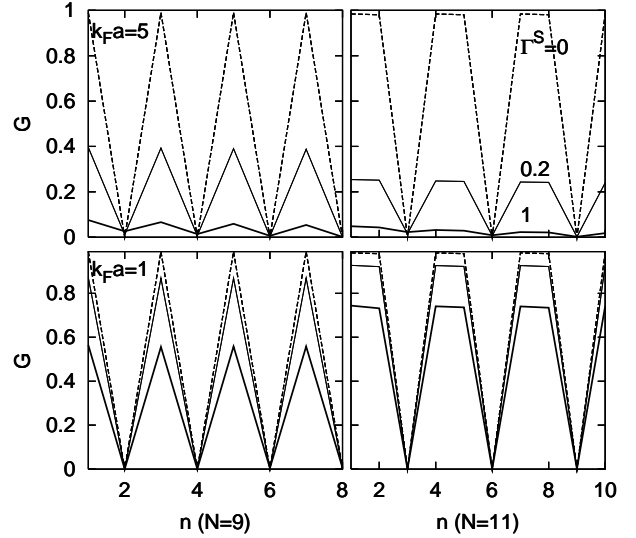}}
\caption{\label{Fig6} The conductance in units [$e^2/h$] as a
function of the STM position $n$, for a wire with $N=9$ atoms and
$\varepsilon_0=0$ (left panels) and for $N=11$ atoms and
$\varepsilon_0=V$ (right panels) for intermediate wavelength such
that $k_Fa=5$ (upper panels) and $k_Fa=1$ (lower panels).  All other
parameters are as in Fig.~\ref{Fig5}. }
\end{figure}

Let us now turn to the more realistic intermediate case of
finite $k_Fa$ and its influence on the conductance oscillations,
Fig.~\ref{Fig6}. Typically this parameter is in the range
$k_Fa=1\ldots10$ \cite{33}.  The onsite energies
are chosen such that the oscillation period is $M=2$ or $M=3$.
The results interpolate those for the limiting cases discussed in
the previous paragraph. Thus, we can conclude that the value of
$k_Fa$ leaves the oscillation decay qualitatively unchanged,
despite the fact that the quantitative difference may be
significant.

\begin{figure}
\centerline{\includegraphics{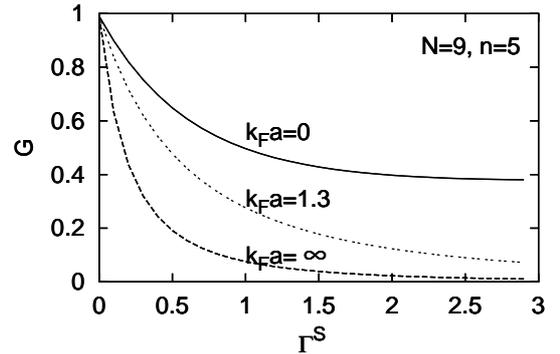}} \caption{\label{fig:gamma}
Decay of the conductance oscillation with the wire-substrate
coupling strength $\Gamma^S$ for a substrate with delocalized
electrons ($k_Fa=\infty$), with localized electrons ($k_Fa=0$) and
for $k_{F} a=1.3$.  The other parameters are $N=9$, $n=5$, and
$\varepsilon_0=0$.}
\end{figure}

In Fig.~\ref{fig:gamma}, we compare the conductance decay for
perfectly localized substrate electrons ($k_Fa=\infty$) and perfectly
delocalized electrons ($k_Fa=0$) with the one for the intermediate value
$k_Fa=1.3$.  One notices that the decay is weaker the more
localized the substrate electrons are.  It is even such that for
$k_Fa=0$, the conductance does not decay entirely, but converges to a
finite value.  This reminds one to the incomplete decay of
entanglement and coherence between delocalized qubits coupled to
substrate phonons \cite{DollEPL,DollPRB}.  A qualitative explanation
for the observed dependence on $k_Fa$ is that an electron that is lost
to a delocalized substrate orbital may tunnel back to its former state
at any site. In the case of a substrate with localized electrons, by
contrast, the lost electron may tunnel back only to the very same wire
site. The rates at which these processes occur should differ roughly
by a factor $N$, i.e., by the number of wire atoms.  This crude
estimate agrees roughly with our numerical results provided that
$\Gamma^L$ and $\Gamma^R$ are of the same order.

\subsection{Charge oscillations}
\label{sec:charge}

\begin{figure}
\centerline{\includegraphics{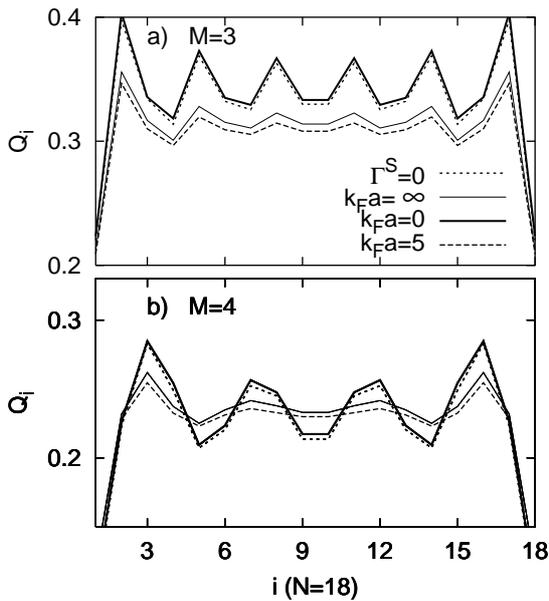}} \caption{\label{Fig7}
Localized onsite charge  $Q_i$ as a function of the atom position
$i$ for a wire consisting of $N=18$ atoms with onsite energies (a)
$\varepsilon_0=V$ and (b) $\varepsilon_0=\sqrt{2}V$. The substrate
Fermi wavelength is $k_Fa=\infty$ (thin solid line), $k_Fa=0$ (thick
solid line), and $k_Fa=5$ (dashed). The dotted lines are obtained in
the absence of the substrate ($\Gamma^S=0$), while the other curves
are the substrate-wire coupling is $\Gamma^S = \Gamma$. The other
parameters are as in Fig.~\ref{Fig2}.}
\end{figure}

We already mentioned that the conductance oscillations relate to
oscillations of the charge density, which can be observed with STM
techniques \cite{34}.  Similar charge waves have been predicted for
a wire with break junction geometry \cite{16,17}.  Thus, one may
suspect that the leakage affects the charge oscillations as well.
Figure~\ref{Fig7} depicts the localized onsite charge
%
%
as a function of the wire atom position $i$ for different surfaces
and the oscillation periods $M=3$ (panel a) and $M=4$ (panel b).
Interestingly, with increasing leakage, the charge oscillations fade
away stronger than the conductance oscillations.  Also here, the
influence of the leakage is more pronounced the more localized the
substrate electrons are. A further common feature is that tunneling
to the substrate affects only the oscillation amplitude, while the
period remains the same.

\section{Conclusions}\label{sec:conclusions}

Using a scattering approach, we have studied conductance and charge
oscillations of a quantum wire in contact with a movable electrode.
As a particular feature, we considered electron leakage from the wire
to various types of substrates.  Our model for the latter allows for
both strongly localized, weakly localized, and perfectly localized
substrate electrons.

As a main feature, we have found that both the conductance
oscillations and the localized onsite charge are fading away due to
the influence of the substrate. Interestingly, this influence is
weaker the more localized the substrate electrons are.  For our model,
the localization parameter $k_Fa$, i.e., Fermi wavelength times the
distance of the wire atoms leads to a monotonic transition between the
limiting cases of perfect localization and delocalization.  In all
cases, the oscillation amplitude is smaller than in the absence of the
substrate.  Thus, leakage represents an obstacle for the experimental
observation of conductance oscillations.  Nevertheless, a piece of
good news is that leakage does not influence the oscillation period.
This is rather encouraging, because it implies that conductance
oscillations can be observed with wires that are grown on surfaces,
despite their unavoidable contact to the wire. We thus are confident
that our results will stimulate STM experiments with wires grown on
vicinal surfaces.

\begin{acknowledgement}
This work has been supported by Grant No.\ N\,N202\,1468\,33 of the
Polish Ministry of Science and Higher Education and by the Alexander
von Humboldt Foundation (TK).  SK and PH acknowledge support by the
DFG via SPP 1243, the excellence cluster ``Nanosystems Initiative
Munich'' (NIM), and the German-Israeli Foundation (GIF, grant no. I
865-43.5/2005 ).
\end{acknowledgement}


\end{document}